\documentclass[sigconf,authorversion,nonacm]{acmart} % AFTER CONDITIONAL ACCEPTANCE

\usepackage{multirow}
\usepackage{enumitem}

\copyrightyear{2023}
\acmYear{2023}
\setcopyright{rightsretained}
\acmConference[DIS Companion '23]{Designing Interactive Systems
Conference}{July 10--14, 2023}{Pittsburgh, PA, USA}
\acmBooktitle{Designing Interactive Systems Conference (DIS Companion '23),
July 10--14, 2023, Pittsburgh, PA, USA}
\acmDOI{10.1145/3563703.3596622}
\acmISBN{978-1-4503-9898-5/23/07}

\begin{document}

%%
%% -------------------------------------------------------------------
%% TITLE
\title{Know What Not To Know: Users' Perception of Abstaining Classifiers}

%%
%% -------------------------------------------------------------------
%% AUTHORS
\author{Andrea Papenmeier}
\email{andrea.papenmeier@uni-due.de}
\affiliation{%
  \institution{University of Duisburg-Essen}
  \streetaddress{Forsthausweg 2}
  \city{Duisburg}
  \country{Germany}
  \postcode{47057}
}
\affiliation{%
  \institution{University of Twente}
  \streetaddress{Drienerlolaan 5}
  \city{Enschede}
  \country{Netherlands}
  \postcode{7522 NB}
}
\author{Daniel Hienert}
\email{daniel.hienert@gesis.org}
\affiliation{%
  \institution{GESIS -- Leibniz Institute for the Social Sciences}
  \streetaddress{Unter Sachsenhausen 6-8}
  \city{Cologne}
  \country{Germany}
  \postcode{50667}
}
\author{Yvonne Kammerer}
\email{kammerer@hdm-stuttgart.de}
\affiliation{%
  \institution{Stuttgart Media University}
  \streetaddress{Nobelstraße 10}
  \city{Stuttgart}
  \country{Germany}
  \postcode{70569}
}
% \author{Christin Seifert}
% \email{christin.seifert@uni-due.de}
% \affiliation{%
%   \institution{University of Duisburg-Essen}
%   \streetaddress{Hufelandstraße 55}
%   \city{Essen}
%   \country{Germany}
%   \postcode{45122}
% }
\author{Christin Seifert}
\email{christin.seifert@uni-due.de}
\affiliation{%
  \institution{University of Duisburg-Essen}
  \city{Essen}
  \country{Germany}
}
\affiliation{%
  \institution{University of Marburg}
  \city{Marburg}
  \country{Germany}
}

\author{Dagmar Kern}
\email{dagmar.kern@gesis.org}
\affiliation{%
  \institution{GESIS -- Leibniz Institute for the Social Sciences}
  \streetaddress{Unter Sachsenhausen 6-8}
  \city{Cologne}
  \country{Germany}
  \postcode{50667}
}

%%
%% SHORT AUTHORS
\renewcommand{\shortauthors}{Papenmeier et al.}

%%
% ----------------------------------------------------- ABSTRACT + TITLE --- 1/2 page
%% ABSTRACT
%% 146 WORDS
%% WIR HABEN EIN !%= WÖRTER LIMIT -WIEVIL? !%= :) 150 :O
\begin{abstract}
  Machine learning systems can help humans to make decisions by providing decision suggestions (i.e., a label for a datapoint). However, individual datapoints do not always provide enough clear evidence to make confident suggestions. Although methods exist that enable systems to identify those datapoints and subsequently abstain from suggesting a label, it remains unclear how users would react to such system behavior. This paper presents first findings from a user study on systems that do or do not abstain from labeling ambiguous datapoints. Our results show that label suggestions on ambiguous datapoints bear a high risk of unconsciously influencing the users' decisions, even toward incorrect ones. Furthermore, participants perceived a system that abstains from labeling uncertain datapoints as equally competent and trustworthy as a system that delivers label suggestions for all datapoints. Consequently, if abstaining does not impair a system's credibility, it can be a useful mechanism to increase decision quality.
\end{abstract}



%%
%% KEYWORDS

%%
%% -------------------------------------------------------------------
%% MAKETITLE
\maketitle
\title{Know What Not To Know}

% ----------------------------------------------------- INTRO
\section{Introduction}

% Motivation
%Today, machines can support humans in a multitude of tasks. 
Decision-support systems (DSS) leverage the computational complexity of machine learning to support users in decision tasks in several domains, e.g., for medical diagnosis~\cite{kaltenhauser2020clinical} or shopping~\cite{herbig2018purchase}. In many use cases, decisions must be made for uncertain or ambiguous datapoints~\cite{bartal1994nuclear,laves2019uncertainty} -- datapoints for which a system might not be able to make a confident decision suggestion. There are automatic methods to detect ambiguous datapoints~\cite{laves2019uncertainty,mol2003neural}. Like humans saying ``I don't know'', the system can make decisions when it is certain but defer uncertain datapoints to a human annotator. In decision-support tasks, however, all datapoints are typically shown to the user alongside the DSS's suggestion for a decision. Users tend to rely on DSS's suggestions, especially for ambiguous datapoints~\cite{papenmeier2022accurate,wang2022accept}, which might lead users toward incorrect decisions. As an alternative, abstaining systems do not deliver a suggestion on highly uncertain datapoints~\cite{ferri2004cautious}. Although methods for equipping a DSS with an abstaining mechanism exist, we do not know how such behavior affects how users perceive the system. This led us to the following research question: \textit{How is the users' perception of a DSS influenced by a system that abstains from offering support on ambiguous datapoints?}

% What we did
To examine this research question, participants of our user study performed a labeling task with the help of a DSS. We varied the DSS's behavior to either abstain or not abstain from suggesting a label for ambiguous datapoints. 
% What we found
Our findings show that users are, often unconsciously, influenced by a system's label suggestion on ambiguous datapoints. An abstaining system does not provide label suggestions on ambiguous datapoints and therefore cannot lead the user toward a wrong decision in those cases. Although an abstaining system explicitly discloses the boundaries of its capabilities to the user, our results suggest that it does not impair perceived system performance or credibility. Our research provides first insights from the users' perspective into abstaining as a mechanism for DSSs to deal with ambiguous datapoints.

% ----------------------------------------------------- RELATED WORK

\section{Related Work}
The ``I don't know'' (IDK) mechanism first appeared in the 1990s in the high-risk domain of anomaly detection in power plants~\cite{bartal1994nuclear}. Usually operating without human interference, these systems delegated uncertain cases to human operators. Contrarily, decision-support systems (DSS) predict a decision (e.g., a label in an annotation task) and suggest the prediction to the user. Mol et al.~\cite{mol2003neural} argued that these systems need a safeguard mechanism to prevent displaying uncertain suggestions. Literature provides several approaches for ``rejecting'' individual datapoints~\cite{gandouz2021machine,laves2019uncertainty}, i.e., classifying them as IDK and subsequently abstaining from labeling the instance. Ambiguous datapoints that fall in the IDK category can be excluded from the dataset during training~\cite{thulasidasan2019combating} to increase dataset consistency or during testing to increase accuracy~\cite{kornaev2020room,trappenberg2000classification}. However, in practice, rejected datapoints still need to be handled, e.g., by being deferred to a human expert~\cite{madras2018predict}, possibly with an explanation of why the sample was rejected~\cite{zhang2022explainable}. However, all works presented above evaluated their algorithms in offline experiments without users. 

%How users perceive a system influences their interaction with an interactive system. 
A DSS helps users make decisions, e.g., in labeling or classification tasks. However, literature shows that people frequently over-rely on DSSs and follow incorrect suggestions, although they would have made a correct decision on their own~\cite{bussone2015,jacobs2021machine}. Especially ambiguous decisions provoke this behavior: Wang et al. found that users' reliance on a DSS increases with reduced confidence in their own decision~\cite{wang2022accept} and Papenmeier et al.~\cite{papenmeier2022accurate} mentioned that users might agree with a system when in doubt. Gandouz et al. argue that an abstaining system, i.e., a system that refrains from showing uncertain suggestions, \textit{``better reflects human decision-making''}~\cite{gandouz2021machine}. Other works similarly call for systems that communicate the boundaries of their capabilities to users~\cite{kawakami2022improving}. A system that uses abstaining to deal with ambiguous datapoints would adhere to this design guideline. Yet, if a system shows high levels of uncertainty, a DSS's perceived credibility (i.e., competence and trustworthiness~\cite{fogg1999elements,winter2014credibility}), which is an important aspect of users' perception~\cite{fogg1999elements}, might be impaired~\cite{bhatt2021uncertainty}. So far, the immediate effects of a system that explicitly and visibly abstains from giving decision suggestions for ambiguous datapoints have not been investigated from the user's perspective.

% ----------------------------------------------------- METHOD 
\section{Method}
%This research investigated the differences between how users perceive a traditional (not abstaining) DSS and how they perceive an abstaining DSS. 
As outlined above, DSSs support users in decision tasks by displaying a suggestion for a decision, e.g., labels in a labeling task. However, ambiguous datapoints are difficult to classify because they provide ambiguous (or not enough) evidence for a clear label decision. The DSS might then be unable to provide reliable suggestions. 
In those cases, an abstaining DSS could reject the sample, i.e., abstain from making a suggestion. To answer our research question, we assessed users' perception of a DSS's performance and credibility after interacting with it in a text labeling task. We focused on the users' perception and used a fictive DSS with simulated output to be able to control its behavior. We employed a between-subjects design and simulated three DSSs, leading to the following conditions:
\begin{enumerate}[itemsep=0mm,nosep]
   \item[\textbf{C1}] \textbf{Correct}: The fictive DSS displays correct label suggestions, i.e., the ground truth, on ambiguous cases.
    \item[\textbf{C2}] \textbf{Abstain}: The fictive DSS does not provide a label suggestion (abstains) on ambiguous cases.
   \item[\textbf{C3}] \textbf{Wrong}: The fictive DSS displays incorrect label suggestions on ambiguous cases.
\end{enumerate}
All three DSSs displayed correct label suggestions for unambiguous cases, which is possible as we control the DSSs' outputs in our setup. The study received clearance from the ethics board of the first author's institution. All materials used in the study (questionnaire, dataset, and anonymized responses) are available online\footnote{\url{https://git.gesis.org/papenmaa/dis23_perceptionofabstaining}}.

\begin{figure*}[t]
    \centering
    \begin{minipage}{.49\textwidth}
      \centering
      \includegraphics[width=0.8\linewidth]{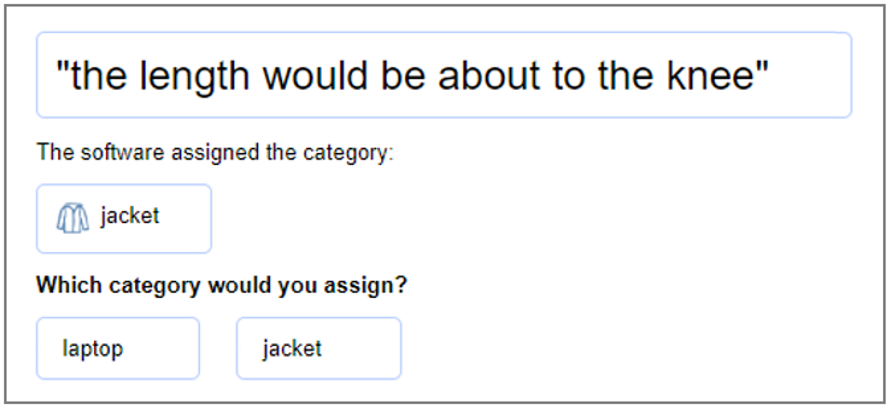}
    \end{minipage}%
    \hfill
    \begin{minipage}{.49\textwidth}
      \centering
      \includegraphics[width=0.8\linewidth]{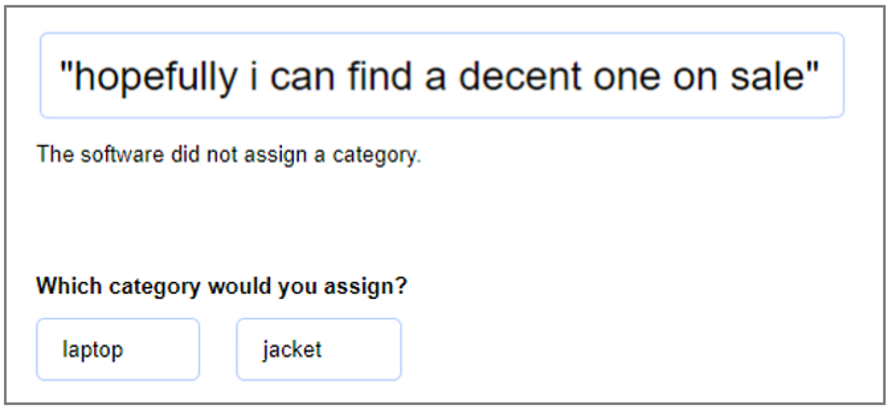}
    \end{minipage}
    \caption{Examples of the task: clear phrase with label suggestion (left) and ambiguous phrase with abstaining system (right).}
     \Description{Two screenshots of the user study task in which participants labeled a phrase as describing either a laptop or a jacket. The left screenshot displays the phrase "the length would be about to the knee" at the top, the fictive DSS's label suggestion "jacket" in the middle, and the two label options "laptop" and "jacket" at the bottom. The label suggestion is introduced with the sentence "The software assigned the category:". The screenshot on the right shows the phrase "hopefully I can find a decent one on sale" with a note that says "the software did not assign a category" in the middle. At the bottom, there are again the two label options "laptop" and "jacket".}
    \label{fig:task}
\end{figure*}

\subsection{Use Case and Dataset}
For analyzing users' perception of a DSS in a labeling task, we needed a dataset with ambiguous datapoints. In their work on human perception of classification mistakes, Papenmeier et al.~\cite{papenmeier2022accurate} published a dataset with 50 phrases that are either ``easy'', ``difficult'', or ``impossible'' to label for human annotators. The phrases were taken from descriptions formulated by users who described either their wishes for a new laptop or a jacket. For example, the jacket phrase \textit{``a budget one that is durable enough to last a long time''} was identified as ``difficult'' to label. At the same time, the ground truth labels were collected along with the descriptions. That is, the ground truth labels were not retrospectively annotated and are therefore not affected by subjective interpretations of annotators. 

To identify those phrases from the dataset that a DSS should abstain from, we asked 30 crowd workers from Prolific\footnote{\url{https://www.prolific.co}} (native English speakers, UK residents, no literacy problems) to annotate the phrases. We offered the label options ``laptop'', ``jacket'', and a residual option ``don't know / unsure''. %During the main phase, two attention checks were added at random places to ensure annotation quality. All crowd workers passed both attention checks. 
We reduced the dataset to 30 phrases to avoid fatigue and boredom effects: Based on the annotations of the crowd workers, we selected 10 laptop and 10 jacket phrases that were unambiguous, i.e., that were correctly annotated by all crowd workers (e.g., \textit{``i would need one with lots of memory and ram available to store large file formats''}). Additionally, we selected 5 laptop and 5 jacket phrases that crowd workers found ambiguous, i.e., that had the highest percentage of ``don't know / unsure'' annotations (e.g., \textit{``quality is most important to me no matter what brand''}, with 63\% of the crowd workers having selected ``don't know / unsure'').

\subsection{Task and Procedure}
First, participants gave informed consent for participation and provided demographic data (age, gender). They then read the scenario that framed the text labeling task as part of a high-quality dataset collection in collaboration with a retail company. The DSS was introduced as a software provided by the retail company. In C2 (abstaining on ambiguous phrases), we introduced the abstaining mechanism as a result of the software not reaching a final decision for some phrases. Participants were then instructed to label the phrases and read the software's suggestions to provide feedback on its performance after the task. Following a training phase with two phrases, participants entered the main phase with 30 phrases (see Figure~\ref{fig:task}) in random order, including two attention checks. Subsequently, participants completed a post-task questionnaire about the perceived performance and credibility of the DSS.

\subsection{Measures and Analysis}
In this experiment, we measured the following three dependent variables:
\begin{enumerate}[itemsep=0mm,nosep,leftmargin=0.3cm]
    \item [] \textbf{User performance}: We calculated participants' labeling accuracy on ambiguous phrases w.r.t. the ground truth. 
    \item [] \textbf{Perceived performance}: Participants rated the DSS's overall performance after the task on a 7-point scale (``1'' = very poor to ``7'' = excellent).
    \item [] \textbf{Perceived credibility}: Participants rated the DSS's credibility after the task on 7-point semantic differentials for competence (unqualified - qualified, inexperienced - experienced, incompetent - competent, Cronbach's $\alpha$ = .91) and trustworthiness (dishonest - honest, untrustworthy - trustworthy, Cronbach's $\alpha$ = .86).% based on~\cite{winter2014credibility}. 
\end{enumerate}
We further gathered qualitative insights into participants' decision-making process: In all three conditions (correct, abstaining, wrong label suggestions), participants also described how the suggestions influenced their labeling behavior. %In \textbf{C2}, participants additionally answered which commonalities they observed in phrases without label suggestions and why they think the system did not give suggestions for those phrases.
We performed one-way ANOVAs with two-tailed Mann-Whitney U-Tests for post-hoc comparisons to compare the three groups regarding the dependent variables. We used a significance level of $\alpha = 0.05$ (two-sided) and report p-values after Bonferroni correction to counteract the repeated testing bias.

\subsection{Participants}
We recruited N~=~120~participants on Prolific (English native speakers, UK residents, no literacy difficulties). All participants received a financial allowance of 1.20~GBP (7.20~GBP/h). Twelve responses were excluded from the analysis due to failed attention checks. The 108 participants with valid responses were, on average, M~=~43.4~years old (SD~=~13.3~years) and approximately balanced between male and female gender (55~female, 52~male, 1~non-binary). Participants in the three conditions did not differ regarding age (F(2,105)~=~0.028, p~=~0.972) or gender distribution ($\chi^2(2)~=~0.050, p~=~0.975$).

% ----------------------------------------------------- RESULTS --- 1 page
\section{Results}
Users frequently adopt incorrect suggestions of a DSS~\cite{bussone2015,jacobs2021machine}, especially on ambiguous datapoints~\cite{papenmeier2022accurate,wang2022accept}. In those cases, an abstaining DSS would lead to better outcomes. To confirm that this behavior is also present in our use case of labeling jacket and laptop sentences, we investigated how a DSS influences users' labeling performance (i.e., how often users chose the correct label). Table~\ref{tab:results} shows the mean performances, i.e., how often participants chose the correct label on average. We define the ``correct'' label as the ground truth label from the dataset. The ANOVA showed a significant main effect. The post-hoc test results showed that providing wrong label suggestions led to a significantly poorer label performance than abstaining (C3 vs. C2: U~=~231.0, p~<~.001) or providing correct label suggestions (C3 vs. C1: U~=~135.5, p~<~.001). The results did not show a difference between C1 and C2 (U~=~476.0, p~=~.078). We also asked participants how the system influenced their decisions. Although the label performance is significantly different in C1 than in C3, many participants reported that the label suggestions did not influence their decisions (47\% in C1, 56\% in C3), e.g., \textit{``It didn't. I used my own judgement of the phrases to decide''} or \textit{``Didn't really influence at all, I went with what I thought about the phras''}.

\begin{table*}[ht]
    \centering
    \caption{Left: Mean user performance on \emph{IDK} phrases in percent with respect to the ground truth. Right: Mean ratings of perceived system performance and credibility and comparison of means with one-way ANOVAs. }
    \label{tab:results}
    \begin{minipage}{.27\textwidth}
        \centering
%    	\label{tab:results_behavior_means}
    	\begin{tabular}{l|rrl} 
            & \multicolumn{3}{c}{\textbf{User}} \\ 
            & \multicolumn{3}{c}{\textbf{Performance}} \\ 
            && M & SD \\
    	    \midrule
    	    \textbf{C1 Correct} && 76\% & 16\% \\ 
    	    \textbf{C2 Abstain} && 69\% & 15\% \\ 
    	    \textbf{C3 Wrong}   && 42\% & 21\% \\ 
            \midrule
            ANOVA & \multicolumn{3}{l}{F(2,105) = 35.13, p < .001}\\
                  %& \multicolumn{3}{l}{p < .001}\\
    	\end{tabular}
        \Description{The table shows the mean labeling performance of participants with respect to the ground truth. In C1 with correct label suggestions, participants had an accuracy of 76\% on average. In C2, where no label suggestions were shown for ambiguous phrases, participants showed an accuracy of 69\%. When seeing incorrect label suggestions on ambiguous data in C3, participants had an accuracy of 42\%. The ANOVA shows a significant difference between conditions with a p-value of under 0.001 .}
    \end{minipage}%
    \hfill
    \begin{minipage}{.69\textwidth}
        \centering
%        \label{tab:results_perceived}
        \begin{tabular}{l|rrl|rrl|rrl} 
            & \multicolumn{3}{c|}{\textbf{Perceived}} & \multicolumn{3}{c|}{\textbf{Perceived}} & \multicolumn{3}{c}{\textbf{Perceived}} \\ 
            & \multicolumn{3}{c|}{\textbf{Performance}} & \multicolumn{3}{c|}{\textbf{Competence}} & \multicolumn{3}{c}{\textbf{Trustworthiness}} \\ 
            && M & SD && M & SD && M & SD \\
    	    \midrule
    	    \textbf{C1 } && 6.00 & 1.11 && 5.75 & 1.22 && 5.78 & 1.33\\ 
    	    \textbf{C2 } && 5.89 & 0.85 && 5.58 & 1.01 && 5.72 & 0.97\\ 
    	    \textbf{C3 }   && 5.75 & 0.79 && 5.75 & 0.96 && 5.88 & 0.89\\ 
            \midrule
            ANOVA & \multicolumn{3}{l|}{F(2,105) = 0.629, p = .535} & \multicolumn{3}{l|}{F(2,105) = 0.301, p = .741} & \multicolumn{3}{l}{F(2,105) = 0.181, p = .834}\\
                  %& \multicolumn{3}{l|}{p = .535} & \multicolumn{3}{l|}{p = .741} & \multicolumn{3}{l}{p = .834}\\
    	\end{tabular}
        \Description{This table lists the mean values of how participants rated the system's performance, its competence, and its trustworthiness per condition. 
        In C1, participants rated the system's performance with 6.00 out of 7 points. C2 has a perceived performance of 5.89 and C3 a perceived performance of 5.75 . The ANOVA does not show a significant main effect with a p-value of 0.535 . 
        For the perceived competence, participants rated C1 with 5.75 points out of 7, C2 with 5.58 out of 7, and C3 with 5.75 out of 7. The ANOVA did not show a significant main effect with a p-value of 0.741 .
        For the perceived trustworthiness, participants rated C1 with 5.78 points out of 7, C2 with 5.72 out of 7, and C3 with 5.88 out of 7. The ANOVA did not show a significant main effect with a p-value of 0.834 .}
     \end{minipage}
\end{table*}

To understand how users perceive an abstaining system, we investigated the perceived performance and perceived credibility (via competence and trustworthiness) in all three conditions. %For the perceived performance, participants rated the DSS's performance on a 7-points scale from ``very poor'' to ``excellent''. 
Table~~\ref{tab:results} (right) presents the mean ratings of the three perception variables. The ANOVA did not show a significant main effect for any of the three variables. %In addition, we measured perceived system credibility via ratings of competence and trustworthiness. Table~\ref{tab:results} shows the mean ratings of competence (middle column) and trustworthiness (right). Similar to perceived performance, the ANOVA did not show a significant main effects in the competence or trustworthiness ratings.

% ----------------------------------------------------- DISCUSSION --- 1 page
\section{Discussion and Conclusion}
In this experiment, we investigated how users perceive different behaviors of decision-support systems (DSS) that either deliver correct (C1), incorrect (C3), or no decision suggestions (C2) for ambiguous datapoints. %Ambiguous datapoints are difficult to classify because they provide ambiguous (or not enough) evidence for a clear label decision. 
Our findings show that participants were strongly influenced by the DSS's label suggestion on ambiguous datapoints and often selected the same label as the system. However, participants were, to a large extent, unaware of this influence as the qualitative data reveals.  

The difference in user performance between the abstaining system (C2) and the correct labeling system (C1) was small (7\% increase) compared to the difference between abstaining (C2) and wrong suggestions (C3) (27\% decrease). That is, in our use case, the risk of steering participants toward wrong decisions was higher than the small gain of providing correct labels. As machine learning systems are likely to perform worse on ambiguous datapoints than on clear datapoints (see~\cite{kornaev2020room,trappenberg2000classification}), abstaining from suggesting a decision for ambiguous datapoints could be advantageous. In the medical domain, for example, a DSS could recommend further tests instead of an uncertain diagnosis when patient data is inconclusive. %Such behavior would avoid misleading the medical professional into an incorrect diagnosis. 
However, future work should investigate why wrong suggestions had a stronger influence on users than correct suggestions.

A potential downside of an abstaining system could be that users perceive a system as less knowledgeable and less trustworthy when it cannot perform the task it was built for (in our case, provide label suggestions). However, our findings do not support this. There was no difference in perceived system performance, perceived competence, or perceived trustworthiness across conditions. As many real-life datasets contain ambiguity, systems need to be prepared to work with ambiguous data. Our findings suggest that abstaining is a potential mechanism to deal with ambiguous datapoints. Therefore, in our future work, we will take a closer look into abstaining mechanisms from the users' perspective and develop design guidelines for abstaining interactive systems. 

In our user study, we used a dataset with clearly ambiguous and clearly unambiguous datapoints and employed fictive DSSs. In reality, datapoints might span the full bandwidth of ambiguity. Practitioners need to identify ambiguous datapoints, for which several methods have been proposed previously \cite{gandouz2021machine,kompa2021second,kornaev2020room,nie2018deeptag}. However, as those methods might introduce additional mistakes and DSSs might make classification mistakes, we want to investigate in the future how users perceive incorrect abstaining behavior, e.g., when the system abstains from labeling clear datapoints. Moreover, we want to test different ways of communicating the reasons for abstaining, e.g., with explanations or the classifier's confidence score.

%%
%% -------------------------------------------------------------------
%% ACKNOWLEDGEMENTS
%\begin{acks}
%asd
%\end{acks}

%%
%% -------------------------------------------------------------------
%% BIBLIOGRAPHY
\bibliographystyle{ACM-Reference-Format}
\bibliography{DIS23PWiP-IDK}

\end{document}